# Transient start-up dynamics and shear banding in aging soft glassy materials: Rate-controlled flow field

Anika Jain,[a] Ramanish Singh,[a] Lakshmi Kushwaha, V. Shankar*, Yogesh M. Joshi*

Department of Chemical Engineering,

Indian Institute of Technology Kanpur, Kanpur 208016. INDIA.

* Corresponding authors, email id: vshankar@iitk.ac.in and joshi@iitk.ac.in

[a] Both authors contributed equally to this manuscript.

**Abstract:**

We study the transient start-up dynamics of a fluidity model which captures the rheological behavior of aging soft glassy materials, in a rectilinear shear flow upon application of step shear rate. We observe that when the steady state flow curve is non-monotonic the system shows transient and/or apparent steady state shear banding in close qualitative agreement with experimental observations. Due to a competetion between aging and rejuvenation during the start up, we show that there is an apparent steady state banding at large times even for shear rates such that the steady state flow curve allows for a homogeneous flow. Thus, for aging soft glassy materials, the shear rate for achieving homogeneous flow is not necessarily given by the steady-state flow curve. We also observe that the transient and apparent steady state shear banding behavior is not correlated to negative slope of the stress-strain dependence during the transient. This work also emphasizes that in order to have a realistic description of shear banding behavior in the aging (time dependent) soft glassy materials, consideration of inertia during the start-up dynamics is crucial.

# I. Introduction:

Many materials of commercial importance such as concentrated suspensions and emulsions, colloidal gels, foams, slurries, and variety of multicomponent systems in pasty form are thixotropic in nature [1,2]. This class of materials, also known as soft glassy materials, undergo physical aging wherein their microstructure evolves as a function of time to attain thermodynamically more stable configurations [3-8]. However, when subjected to a deformation field, a soft glassy material rejuvenates wherein the structure evolved due to physical aging reverts. The rheological behavior of a material is therefore is an outcome of the competition between dominance of aging and rejuvenation for a given strength of deformation field and microstructure [9-12]. By virtue of thixotropy some of the soft glassy materials show yield stress, a threshold stress below which material undergoes no/slow flow [13,14]. Interestingly, in a drag flow in rectilinear as well as curvilinear Couette geometries, this class of materials have been observed to demonstrate apparent steady state as well as transient shear banding wherein the flow field gets divided into bands of distinctly different shear rates [14,15]. In this work we analyze the banding phenomenon by solving the transient dynamics as exhibited by a fluidity model in simple shear flow by accounting for inertial effects in the material.

Soft glassy materials are composed of constituents that are structurally arrested by their neighbors either due to physical jamming (owing to crowding of the same) or because of being trapped in energy wells (as a result of interparticle interactions). Such structural arrest inhibits mobility of the constituents, and therefore kinetically constrains the system from attaining the thermodynamic equilibrium state. By virtue of the constituents being trapped in different kinds of cages (real or energetic), there exists a distribution of relaxation modes associated with soft glassy materials, and owing to the microscopic dynamics that every constituent undergoes, while undergoing physical aging, these relaxation modes slow down as a function of time causing evolution of the whole distribution [3]. The rheological behavior of a system as a whole is determined by how the individual modes undergo aging and rejuvenation under application of a particular kind of deformation field leading to alteration of the relaxation time distribution as a function of time [16,17]. Such competition between aging modes in comparison to rejuvenation modes are proposed to render the soft glassy materials to exhibit a rich array of rheological behavior such as yield stress [14,18-20], viscosity bifurcation [13,21], delayed yielding [22,23], delayed solidification [24-26], shear banding [14,15], overaging [11,27], etc.

The presence of yield stress that marks transition from a no/slow flow to flow in a material very naturally leads to shear banding in a pressure-driven flow where there exists shear localization or a gradient of shear stress [28]. However, in case of a rectilinear Couette flow, where stress is uniform across the gap or for curvilinear Couette flow with a thin gap where stress gradient is very weak, a conventional yield stress material that follows either Bingham or Herschel–Bulkley constitutive equation shows a homogeneous flow above the yield stress. These phenomena are well understood and are part of the pedagogical literature. In the recent literature of past two decades, there have been number of reports wherein soft glassy materials have been observed to show transient as well as steady state shear banding even in simple shear (Couette) flow [14,15]. In a typical observation, for different types of soft glassy materials, a number of groups noted that when a material undergoing shear flow at high shear rate, wherein it shows a homogeneous flow field, is subjected to a step-down shear rate jump to a shear rate below a certain critical value, an apparent steady state shear banding is observed [15,29-32]. In this case flow field gets divided into two parts with a stationary band near the fixed boundary while the flowing band is near the moving boundary. Interestingly the flowing band is always observed to possess a constant shear rate that is equal to the critical value. With increase in the imposed shear rate, the width of the flowing band is observed to increase according to a lever rule such that for imposed shear rate equal to that of the critical value a homogeneous flow is observed [8,31]. It was proposed that the observed behavior originates from a non-monotonic flow curve with an unstable branch (shear stress – shear rate curve having a negative slope) below the critical shear rate [13,33].

Rather than analyzing the velocity profile upon step down jump in shear rate, analogous behavior has also been studied by allowing a soft glassy material to age under quiescent conditions, before applying shear rate step jump. Martin and Hu [34] studied aqueous suspension of Laponite, wherein they reported three different kinds of behavior depending upon an applied shear rate. Firstly, irrespective of the magnitude of shear rate, soon after applying a step shear rate, an aged Laponite suspension showed transient shear banding with the band next to fixed boundary being stationary. For a lower shear rate, the width of stationary band increased with time finally showing a steady state shear band. For moderate shear rates, the width of stationary band shrunk with time, eventually leading to a steady state shear band. For sufficiently high shear rates also the width of the stationary band shrunk with time, however, in due course vanished bringing about a homogeneous shear flow. It was further observed by Martin and Hu [34] that an aqueous suspension of Laponite shows a non-monotonic flow curve and they attributed the observed behavior to a competition between aging and rejuvenation in the flowing and stationary bands [34]. Kurokawa and coworkers [32] performed step shear rate experiments on aged Ludox gel. Subsequent to shear melting, for lower waiting times, they observe that Ludox gel did not show transient as well as steady state shear banding. For moderately high waiting times, however, Ludox gel did show transient shear banding that eventually transformed to homogeneous flow. On the other hand, for high enough waiting times they observed transient as well as steady state shear banding. They also changed the magnitude of shear rate, wherein they observed steady state shear banding for low shear rates while transient shear banding evolving to homogeneous flow for the high shear rates. They also reported noticeable slip at the wall. Rogers and coworkers [35] studied multi-arm star polymers system that shows yield stress plateau if the shear rate is decreased very slowly from a high value. Equivalently, the system is observed to show a strain plateau (solidification) after a prolonged creep at stress just below the yield stress. In a shear start up experiment the system shows absence of banding at smaller times elapsed since shear melting. However, after a prolonged deformation the system shows banding. Rogers and coworkers [35] analyze the rheological behavior using fluidity model.

Similar to experimental work on shear banding in aging systems, non-aging or very weakly aging yield stress materials have also been studied that usually show only a transient banding. Divoux *et al.* [36,37] studied Carbopol gel under controlled shear rate conditions and observed that the system remains in a transient shear banded state for a prolonged time before suddenly acquiring a homogeneous state. The time taken by the system to achieve a homogeneous state is observed to show an inverse power law dependence on shear rate. Along similar lines, Divoux *et al.* [38] observed transient shear banding for a stress controlled flow field, wherein time to achieve a homogeneous state showed an inverse power law relation on stress in excess to yield stress. Grenard and coworkers [39] studied rheological behavior of carbon black suspension and reported transient shear banding for the same. They observed that time required to acquire the homogeneous state depends exponentially on stress. Ovarlez and coworkers have analyzed the origin of flow inhomogeneities during the transient shown by non-aging or very weakly aging materials [40].

There have been many theoretical efforts to capture the rheological behavior of soft glassy materials. The key to all the modeling efforts is to understand how individual relaxation modes in a spectrum age with time and get affected by the deformation field resulting in possible movement and alteration of a spectrum. The most rigorous progress in this endeavor is the development of soft glassy rheology model by Cates and coworkers [3,41]. They consider a model soft glassy material with large number of mesoscopic entities arrested in energy wells with depths following Boltzmann distribution. For thermal noise below that associated with the point of glass transition, the Boltzmann distribution is not normalizable. Consequently, the depth of the energy wells go on increasing thereby naturally leading to physical aging. They consider escape of the trapped entities from the respective wells as thermally or strain activated. Upon escape an entity chose a well randomly from a preassigned distribution leading to profound alteration of well depth distribution. The success of SGR model is in its representation of relaxation time spectrum and how this gets translated and modified with age and deformation field in a self-consistent manner. In addition to many usual signatures of the experimental behavior of aging soft glassy materials, the SGR model successfully predicts more subtler effects such as overaging [11,17,27]. The original SGR model predicts the steady state stress – shear rate relationship to be monotonic. Fielding *et al.* [42] extended the SGR model to include evolution of noise leading to non-monotonic steady state stress – shear rate relationship broadening the predictive capacity of the model.

Many of the experimental observations of soft glassy materials are qualitatively predicted by 'toy' fluidity models, which are minimal models that encapsulate the key rheological observations. This class of models are obtained by expressing a rate equation for structure parameter (usually represented by $\lambda$) that represents the extent of structure formed during the physical aging. In the rate equation, the time derivative of $\lambda$ is expressed as: $d\lambda/dt = A(\lambda,\dot{\gamma}) - R(\lambda,\dot{\gamma})$, wherein the aging $(A)$ and rejuvenating $(R)$ contributions are usually considered to be dependent on timescales and/or strength of deformation field. An exhaustive list of various functional forms for $A$ and $R$ can be found elsewhere [2,43]. Coussot [4] proposed the following equation for $\lambda$ evolution, which along with a term representing diffusion of structure is given by:

$$\frac{\partial \lambda}{\partial t} = \frac{1}{T_0} - R(\lambda,\dot{\gamma}) + D\frac{\partial^2 \lambda}{\partial y^2}. \tag{1}$$

Since the deformation field is primarily responsible only for rejuvenation, and the parameter $\lambda$ does not represent any physically measurable quantity in soft glassy materials, equation (1) represents the most general form for $\lambda$ evolution. The first term on the right suggests that aging of parameter $\lambda$ occurs with a constant timescale $T_0$: smaller a value of $T_0$ is, faster is aging. In the third term, $y$ is the direction of velocity gradient, while the parameter $D$ is diffusivity. This term suggests fluidity or rigidity at any region gets influenced by that of the surrounding region. This diffusion term is particularly relevant when fluidity model is to be solved over a specific geometry. In addition to evolution of fluidity, the model also requires a constitutive relation and a relation between parameters of the constitutive relation such as viscosity, modulus, etc., and $\lambda$. Owing to empiricism associated with relating $\lambda$ and the parameters of the constitutive relation, the fluidity models, depending upon choice of parameters, show monotonic as well as non-monotonic steady state shear stress – shear rate relationships, thereby imparting significant flexibility to the model. However, usually the fluidity models are single mode models and therefore cannot predict those effects that are caused by alteration of a relaxation time distribution.

A considerable amount of theoretical work has been carried out to analyze shear banding in aging systems. Moorcroft *et al.* [44] studied the transient shear banding in the soft glassy materials using two different theoretical models. The first model is a simple fluidity model coupled with a Maxwell model to account for the elastic glassy degrees of freedom in a material. The second is an adapted version of the soft glassy rheology model. Their fluidity model assumed instantaneous momentum diffusion (neglecting inertia of the material), and the total stress in the model is given by a sum of the viscoelastic stress (from the glassy degrees of freedom) and a pure viscous part. The viscoelastic stress was governed by a simple Maxwell model, but with a time-dependent relaxation time (whose inverse is a measure of fluidity). Under steady-state conditions, their model shows a stress plateau in a limit of small shear rates associated with a constant a yield stress. The dynamics of the relaxation time was governed by an aging term, a rejuvenation term proportional to the relaxation time itself, and a diffusive term that represented the tendency of the relaxation time to equalize with that of adjacent regions. Under the application of a step increase in shear rate, their constitutive model showed a nonmonotonic behavior for stress versus strain (or, equivalently, time). After an initial increase in the stress due to the elastic effects in the model, eventually, the stress decreased to approach the steady state value. Furthermore, the magnitude of the stress overshoot increased with increase in waiting time before the application of the step increase in the shear rate. They carried out an `instantaneous' linear stability analysis when the stress in the system started to decrease from its maximum value, and showed that the system is unstable in this decreasing part of the stress-strain curve. They further carried out numerical simulations of the step change in shear rate, by allowing for transverse variations in the flow-gradient direction. Crucially, they imposed an inhomogeneous, $y$-dependent stress profile as an initial condition to initiate the banding process. Their numerical results indicated that regions of maximum shear banding (as measured by the difference in the shear rates of the two bands) agreed with the regions of negative slope in the stress-strain curve. Interestingly, their velocity profiles in the banded state showed *negative* velocities and shear rates in the low-shear rate band, which was attributed to the elastic nature of the low shear band due to aging. In a subsequent work, Moorcroft and Fielding [45] carried out a linear stability analysis to arrive at generic criteria for transient shear banding for various flow protocols such as a step change in stress or shear rate. They elaborated this by using the Rolie-Polie model for polymers and the soft-glassy rheology model. For a step increase in shear rate, they concluded that if the time derivative of stress in the system is negative, then the system should display transient banding. More recently, Radhakrishnan *et al.* [46] followed up the work of Moorcroft *et al.* [44] by extending the theory to viscosity bifurcating yield stress fluids. However, the other assumptions of the earlier work of Moorcroft *et al.* [44] such as no inertia and inhomogeneous initial stress condition above were adpoted in Radhakrishnan *et al.* [46] as well. Their simulations successfully captured the rheologial hystersis as observed in experiments.

While these prior theoretical efforts to understand and predict transient banding have neglected inertia, there have been some studies [47] in the past using models for worm-like micellar systems that have argued that inertial effects play a very important role in the transient dynamics and subsequent banding. Zhou *et al.* [47], showed that the coupling of inertia with elasticity in their model resulted in the formation of damped inertio-elastic waves [48,49], which caused the formation of multiple transient shear bands during the start up flow. It is important to recognize that the work of Zhou *et al.* has elasticity in their constitutive model. The phenomena predicted by Zhou *et al.* [47] were directly a consequence of fluid inertial effects albeit in the context of a non-aging fluid.

The question that arises naturally is what are the essential ingredients for transient banding, especially in the context of aging soft-glassy materials: could transient banding arise in such systems even in the absence of elasticity? In this context, recently, Illa and co-workers [50] used a simple time-dependent (inelastic) Newtonian fluid, and ignored other features of yield stress fluids and thixotropic fluids, such as elasticity and yield stresses. They coupled the rheological model to the Cauchy momentum equation as applicable for a concentric cylinder Couette device. Transient shear banding occurs in their formulation due coupling between the shear rate and viscosity in the Navier-Stokes solution. Unlike many other previous studies, the authors included inertial effects in the fluid. However, the model used by them for the evolution of the structure ignores aging and concomitant viscosity build up in the absence of flow. In their model, the growth of structure itself is proportional to the prevalent shear rate in the fluid. This is a major point of departure from the present study that models a spontaneously aging system under quiescent conditions, which show shear banding upon imposition of step shear rate. The recent work of Korhonen and co-workers [51] extended the earlier work of Illa *et al.* to the plane Couette geometry. In both their studies, the flow achieved a homogeneous steady state at sufficiently large times.

In this work, we analyze the transient dynamics under start up for a fluidity model that shows montotonic as well as the non-monotonic steady state flow curve in the plane Couette geometry by including inertia. Our work, on one hand, for the first time qualitatively predicts different experimental observations related to transient banding in soft-glassy fluids reported in the literature. On the other hand, we critically analyse the modelling efforts in this field, in particular, the necessity of including inertia and the requirement of elasticity in a constitutive relation while solving fluidity models for aging soft glassy materials to analyse the transient as well as the apparent steady state shear banding. We also assess the results of our study with respect to the rheological signatures that have been proposed in the literature to indicate shear banding. The rest of this paper is organized as follows: We provide a description of the model used and the governing equations in Section II. Section III presents the results obtained from our numerical study and the connection to existing experimental results. In the same section we also discuss how the present results relate to or differ from the existing theoretical efforts. Finally, we conclude with the salient conclusions of this work in Section IV.

**II. Model and Governing Equations**

In this work, we use the fluidity model proposed by Coussot *et al.* [13] where the evolution of structure parameter $\lambda$ (which is a measure of the extent of structure formation during aging) is given by,

$$\frac{d\lambda}{dt} = \frac{1}{T_0} - \beta\lambda\dot{\gamma}, \tag{2}$$

where $\beta$ is a model parameter. Equation (2) suggests that the rate of $\lambda$ evolution is proportional to reciprocal of constant time scale $T_0$. On the other hand, the rate of rejuvenation is proportional to the product of instantaneous value of $\lambda$ and the second invariant of rate of strain tensor given by: $\dot{\gamma} = \sqrt{\dot{\underline{\gamma}}:\dot{\underline{\gamma}}/2}$, where $\dot{\underline{\gamma}}$ is the rate of strain tensor. Furthermore, the constitutive relation for relating the stress tensor $\left(\underline{\sigma}\right)$ and $\dot{\underline{\gamma}}$ is considered to be the generalized Newtonian model [28] and is given by:

$$\underline{\sigma} = \eta \dot{\underline{\gamma}}, \tag{3}$$

where $\eta$ is viscosity. It is proposed that $\eta$ explicitly depends only on $\lambda$, which in turn depends on time and shear rate through equation (2), and is given by:

$$\eta = \eta_0 f(\lambda) = \eta_0 \left(1 + \lambda^n\right). \tag{4}$$

Here $\eta_0$ is the viscosity associated with the completely rejuvenated state $(\lambda = 0)$ of a soft glassy material, while the power law exponent $n$ determines the strength of the enhancement of viscosity. Equations (2) to (4) completely define the fluidity model.

In this work, we analyze start-up flow of the above-mentioned fluidity model in a simple (rectilinear) shear flow geometry as shown in figure 1. In a typical flow protocol, a completely rejuvenated material is allowed to age for time $t_w$ under quiescent conditions before the top plate is set in motion with constant velocity $U$ in the $y$ direction. The bottom plate, which is at a distance $H$ beneath the top plate in the $y$ direction, is always stationary leading to gradient in velocity in the $y$ direction. The $z$ direction is the vorticity direction. For the specified kinematics of the velocity field, the Cauchy momentum equation is given by:

$$\rho \frac{\partial u_x}{\partial t} = \frac{\partial \sigma_{xy}}{\partial y}, \tag{5}$$

where $\rho$ is density, $u_x$ is the $x$ component of the velocity vector and $\sigma_{xy}$ is the $xy$ component of the stress tensor. For a simple shear flow of a generalized Newtonian model, only the $x$ component of the velocity vector ($u_x$) and $xy$ and $yx$ components of the stress tensor ($\sigma_{xy}$ and $\sigma_{yx}$ respectively) are nonzero. Furthermore, in the rate of strain tensor only $\partial u_x / \partial y$ is non-zero. Since for a unidirectional simple shear flow of a incompressible material mass balance is automatically satisfied, simultaneous solution of equations (2) to (5) leads to complete description of the flow field.

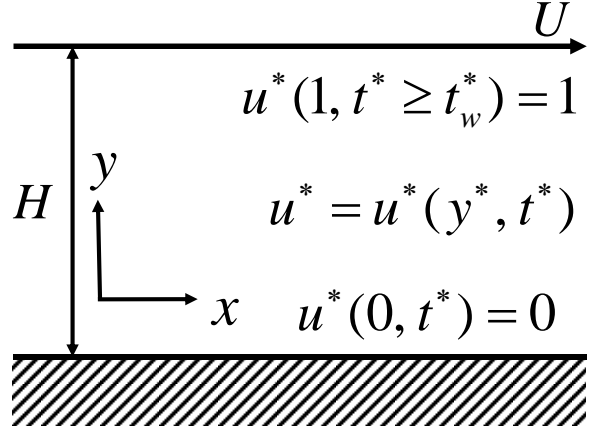

**Figure 1.** Planer Couette geometry with the bottom plate to be stationary. In this work the top plate is moved at velocity $U$ (step shear rate) at time $t_w^*$ after rejuvenation.

In order to simplify the set of equations to be solved, we use the following dimensionless parameters and equations. Velocity $u_x$ is non-dimensionalized by the top plate velocity, $U$, thus $u^* = u_x/U$. Time is non-dimensionalized using kinematic viscosity, $\eta_0/\rho$, and the gap between the plates, $H$, leading to: $t^* = \eta_0 t/\rho H^2$. The $y$ coordinate is non-dimensionalized as $y^* = y/H$, while the dimensionless stress is represented by: $\sigma^* = \sigma_{xy}\beta T_0/\eta_0$. Using these dimensionless parameters, the relationship between stress and strain rate is given by:

$$\sigma^* = \left(1+\lambda^n\right)\frac{\partial u^*}{\partial y^*}\dot{\gamma}_0^*, \tag{6}$$

where $\dot{\gamma}_0^*$ is the dimensionless applied global shear rate given by: $\dot{\gamma}_0^* = \dfrac{\beta T_0 U}{H}$. Furthermore, the non-dimensional evolution equation for $\lambda$ is given by:

$$\alpha\left(\frac{\partial \lambda}{\partial t^*}\right) = 1 - \lambda\dot{\gamma}_0^*\left(\frac{\partial u^*}{\partial y^*}\right), \tag{7}$$

where $\alpha = T_0\eta_0/\rho H^2$ is the dimensionless number representing ratio of aging time $T_0$ to momentum diffusion timescale $\rho H^2/\eta_0$. When $\alpha < 1$ aging and momentum diffusion both are important and can influence the the transient evolution of such a fluid. Previous studies [35,44-46,52], which have implicitly considered infinitely fast momentum diffusion would correspond to the limit $\alpha \to \infty$. Finally, the dimensionless momentum balance can be represented as:

$$\frac{\partial u^*}{\partial t^*} = \frac{1}{\dot{\gamma}_0^*}\frac{\partial \sigma^*}{\partial y^*}. \tag{8}$$

The equations (6) to (8) are solved subject to following initial and boundary conditions. At time $t^* = 0$, the material is considered to be in a completely rejuvenated state. Experimentally such a state is achieved by shear melting a sample, leading to:

$$\lambda = 0 \text{ at } t^* = 0 \text{ for all } y^*. \tag{9}$$

Thereafter the material is kept under quiescent conditions until time $t_w^*$. During this time, it ages according to equation (7) with $\dot{\gamma}_0^* = 0$ resulting in following condition (during which $\lambda$ evolves as, $\lambda(t^*) = t^*/\alpha$):

$$u^* = 0 \text{ for } t^* < t_w^* \text{ at all } y^*. \tag{10}$$

Finally, for $t^* \geq t_w^*$ the top plate is moved with a constant velocity:

$$u^* = 1 \text{ at } y^* = 1 \text{ for } t^* \geq t_w^*, \tag{11}$$

and no-slip boundary condition on both the plates imply:

$$u^* = 0 \text{ at } y^* = 0 \text{ for } t^* \geq t_w^*. \tag{12}$$

We solve equations (6) to (8) subject to above conditions in a partial differential equation solver in COMSOL 4.3®. We also carry out some of the simulations in Matlab® to validate the results obtained using COMSOL 4.3®. We also check effect of the mesh size on the results. We observe that below a mesh size $5\times10^{-3}$ the results are independent of the mesh size.

**III. Results and Discussion:**

We first discuss the steady state shear stress – shear rate relationship (i.e. the steady state flow curve) of the model as described by equations (6) and (7) that assumes homogeneous steady velocity field given by: $u^* = y^*$. Setting the time derivative in equation (7) to zero, the steady state shear stress and shear rate are given by:

$$\sigma_{ss}^* = \left(1+\lambda_{ss}^n\right)/\lambda_{ss} \text{ and } \dot{\gamma}_{ss}^* = 1/\lambda_{ss}, \tag{13}$$

where $\lambda_{ss}$ is the steady state value of $\lambda$. In figure 2 we plot $\sigma_{ss}^*$ as a function of $\dot{\gamma}_{ss}^*$ for three values of $n = 0.7$, 1 and 1.2. It can be seen that for $n = 0.7$, the model predicts a monotonic flow curve. This curve, while belonging to a thixotropic material, leads to a stable value of shear stress for any applied shear rate. Equation (13) suggests that similar monotonic flow curves exist for $n < 1$. For $n = 1$, the model predicts a stress plateau in the limit of low shear rates leading to a constant yield stress thixotropic fluid, while for $n = 1.2$, the model predicts a non-monotonic flow curve. According to Equation (13), the non-monotonic flow curve exists for $n > 1$, wherein dependence of shear stress ($\sigma_c^*$), shear rate ($\dot{\gamma}_c^*$) and $\lambda$ associated with the minimum on $n$ is given by:

$$\sigma_c^* = n(n-1)^{(1-n)/n}, \ \dot{\gamma}_c^* = (n-1)^{1/n}, \text{ and } \lambda_c = (n-1)^{-1/n} \qquad \text{for } n>1. \qquad (14)$$

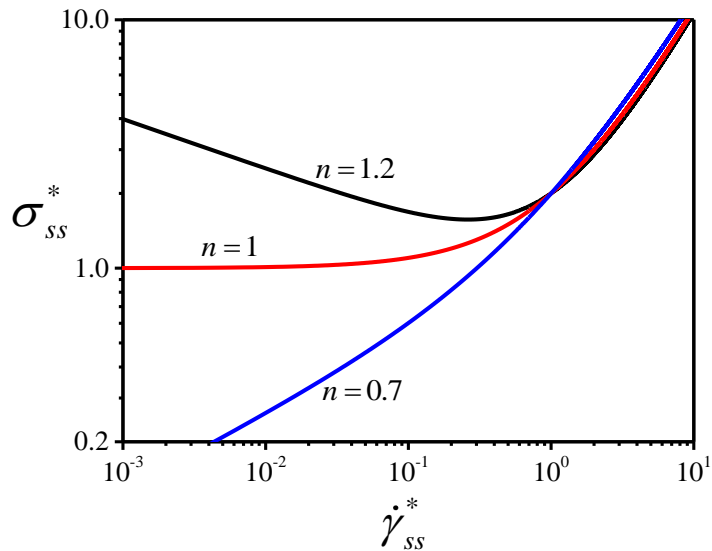

**Figure 2.** The steady state shear stress – shear rate constitutive curve obtained by solving equation (13). A flow curve becomes non-monotonic for $n > 1$.

Figure 2 shows that, for shear rates lesser than $\dot{\gamma}_c^*$, steady state stress is a decreasing function of the shear rate, which renders it linearly unstable. Very importantly, the non-monotonic flow curve suggests that in a spontaneously aging material under quiescent conditions, yield stress remains constant at $\sigma_c^*$ over the initial period until $\lambda$ remains smaller than $\lambda_c$. During subsequent aging, as $\lambda$ increases beyond $\lambda_c$, yield stress increases as a function of time. The dependence of $\sigma_c^*$ on $\lambda$ is given by:

$$\sigma_y^* = \sigma_c^* \text{ for } \lambda \leq \lambda_c \text{ and } n > 1$$

$$\sigma_y^* = \left(1 + \lambda^n\right)/\lambda \text{ for } \lambda > \lambda_c \text{ and } n > 1 \tag{15}$$

A detailed discussion on how the non-monotonic flow curve as shown in figure 2 leads to time dependent yield stress can be found elsewhere [33].

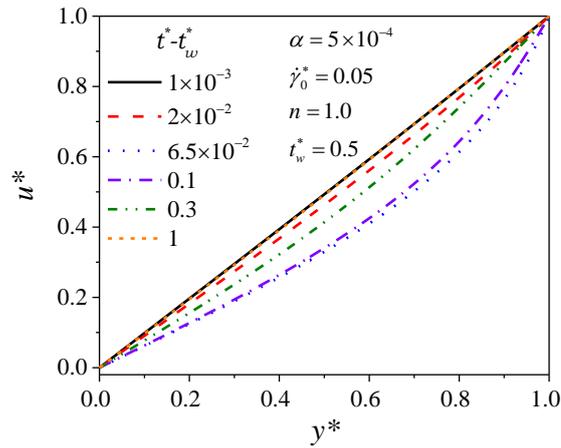

(a)

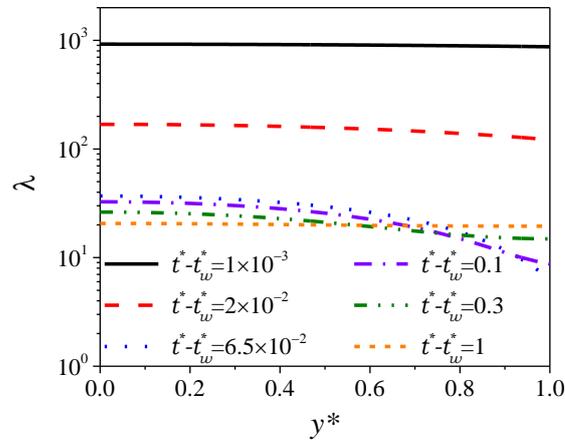

(b)

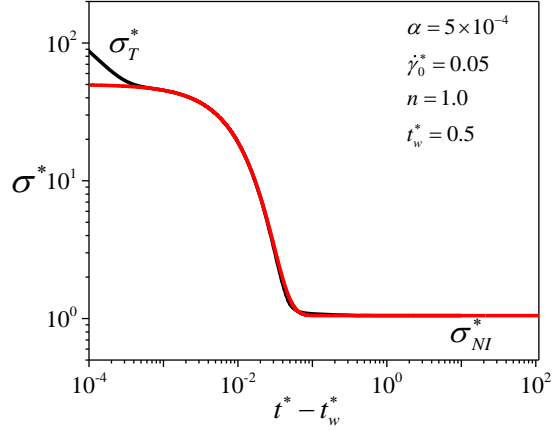

(c)

**Figure 3.** Velocity profiles (a), $\lambda$ profiles (b), and $\sigma_T^*$ and $\sigma_{NI}^*$ (c) at different $t^* - t_w^*$, where $t_w^* = 0.5$ for $n = 1$, $\alpha = 5 \times 10^{-4}$ under application of $\dot{\gamma}_0^* = 0.05$. It can be seen that the velocity profile in a limit of small $t^* - t_w^*$ is near homogeneous. With increase in $t^* - t_w^*$ the profile deviates from the homogeneous profile without showing any transient shear banding. In a limit of high $t^* - t_w^*$ the profile again becomes homogeneous. On the other hand, both $\sigma_T^*$ and $\sigma_{NI}^*$ show decrease with $t^* - t_w^*$ (or equivalently the strain) leading to a time invariant value.

As discussed above, the solution of equation (13) clearly suggests that for $n \leq 1$, the flow curve is monotonic and stable for all the values of applied strain rate. As a result, even though a plateau of constant stress is observed for $n = 1$, the response to strain rate controlled flow field for $n < 1$ is qualitatively similar to that of for $n = 1$. Therefore we discuss the results associated with only the case of $n = 1$. In figure 3(a) we plot velocity profiles at different $t^* - t_w^*$ for $n = 1$, $\alpha = 5 \times 10^{-4}$ and $t_w^* = 0.5$. In this case a step shear rate is applied at time $t_w^*$ after shear melting is stopped. During the waiting period, material ages according to equation (7) except the last term. As a result viscosity of the material at the time when step shear rate is applied is significantly higher. It can be seen that at very short time of $t^* - t_w^* = 10^{-3}$, the velocity profile is near homogeneous, which arises due to high viscosity evident from high value of $\lambda$ shown in figure 3(b) causing rapid momentum diffusion. Subsequently, the velocity profile gradually deviates from the homogeneous profile showing greater gradient near the moving wall. However, as time increases, the effect of last term in equation (7) becomes apparent and material rejuvenates leading to overall decrease in $\lambda$, but with greater intensity near the moving wall. Eventually the rejuvenation near the stationary plate and that of near the moving plate approach the same constant value bringing about homogenization of the velocity profile. In the limit of large times a completely homogeneous profile is obtained, however during this entire evolution no clear signatures of transient shear banding is observed. In figure 3(c) we plot $\sigma_T^*$ ($\sigma^*$ at the top plate $y^*=1$) as a function of $t^*$. Initially the stress is high due to acceleration of the fluid during the transient phase. As time increases, $\sigma_T^*$ decreases and attains a constant value as the steady state is reached.

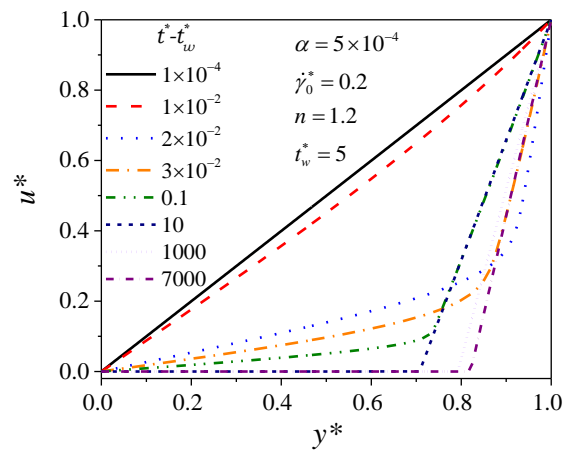

(a)

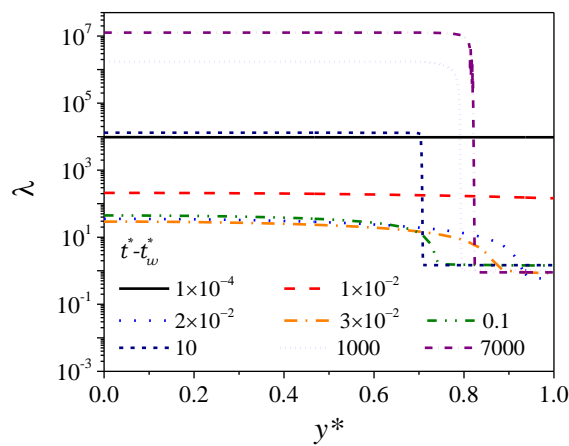

(b)

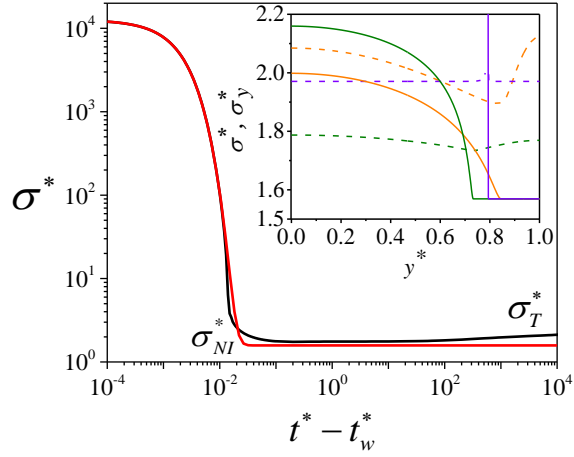

(c)

**Figure 4.** Velocity profiles (a), $\lambda$ profiles (b), and $\sigma_T^*$ and $\sigma_{NI}^*$ (c) at different $t^*-t_w^*$, where $t_w^*=5$ for $n=1.2$, $\alpha=5\times10^{-4}$ under application of $\dot{\gamma}_0^*=0.2<\dot{\gamma}_c^*$. It can be seen that the velocity profile in a limit of small $t^*-t_w^*$ is near homogeneous. With increase in $t^*-t_w^*$ the profile deviates from the homogeneous profile leading to transient shear banding. With further increase in $t^*-t_w^*$ the width of stationary band increases eventually leading to apparent steady state banding. In figure (c), both $\sigma_T^*$ and $\sigma_{NI}^*$ show decrease with $t^*-t_w^*$ (or equivalently the strain). The former, however starts increasing at higher times while the latter shows a time invariant value. Inset shows stress (dashed) and yield stress (solid) profiles for time $t^*-t_w^* = 0.03$ (orange), 0.1 (green) and 1000 (purple).

Next we discuss the case of $n=1.2$ for which the flow curve shows a non-monotonic behavior as shown in figure 1. According to equation (14), for $n=1.2$, $\sigma_c^*=1.57$, $\dot{\gamma}_c^*=0.26$ and $\lambda_c=3.8$. For this case, the yield stress depends on $\lambda$, and is given by equation (15). As discussed before, the non-monotonic flow curve presents a very peculiar case for which the branch of flow curve with negative slope is always linearly unstable, and therefore it is not possible to sustain a homogeneous steady flow for shear rates in the unstable region. We first discuss the case wherein the imposed shear rate is in the unstable part of the flow curve. In figure 4(a) we plot the velocity profile for this system at different $t^*-t_w^*$ for an imposed $\dot{\gamma}_0^*=0.2$ applied at $t_w^*=5$ for $\alpha=5\times10^{-4}$. In figure 4(b) we plot $\lambda$ while in the inset of figure 4(c) we plot $\sigma^*$ and $\sigma_y^*$ as a function of $y^*$ at diffetent values of $t^*-t_w^*$. It can be seen that at very small time $t^* - t_w^* = 10^{-4}$ the velocity profile is near homogeneous. This is because for $t_w^* = 5$, the value of $\lambda$ is very high that results in very high value of viscosity (momentum diffusivity); consequently the velocity profile develops at small value of at $t^* - t_w^* = 10^{-4}$. Although the velocity profile is almost homogeneous, the strain induced within the sample at $t^* - t_w^* = 10^{-4}$ is extremely small: $\dot{\gamma}_0^* (t^* - t_w^*) = 2 \times 10^{-5}$. The reason for mentioning velocity profile to be *almost* homogeneous as there always exists a very small gradient in shear rate with slightly greater value near the moving wall. As the time progresses, the material starts rejuvenating as per equation (2) (equation (7) in dimensionless form) with greater extent near the moving wall leading to decrease in $\lambda$ and $\sigma_y^*$ as observed for $t^* - t_w^* = 10^{-2}$. As a result, at this time the deviation in velocity profile from the homogeneous flow, although slight, becomes apparent. This further enhances rejuvenation near the moving wall compared to that of at the stationary wall. The inset in figure 4(c) shows that for $t^* - t_w^* = 0.03$, $\sigma^*$ induced in material is greater than the $\sigma_y^*$ (the same behavior is observed for $t^* - t_w^* \leq 0.03$), however as time progresses to $t^* - t_w^* = 0.1$, while $\sigma_y^*$ of material near the moving wall remains constant, $\sigma_y^*$ of material near the straitionary crosses $\sigma^*$ in the material at that location. Consequently, at $t^* - t_w^* = 0.1$, material is significantly rejuvenated near the wall than away from it showing a noticeable transition at $y^* \approx 0.7$. This results in two bands, a fast flowing band near the moving wall while a slow flowing band away from the wall with transition at $y^* \approx 0.7$ (in the present model framework $\sigma^* < \sigma_y^*$ does not mean the flow stops, such scenario simply means $\lambda$ must increase indefinitely without showing a steady state thereby leading to continuous decrease in shear rate). At larger times, the material near the moving wall continues to rejuvenate but only with a marginal decrease in $\lambda$. However, since in this region $\lambda < \lambda_c$, $\sigma_y^*$ remains constant at a value associated with minimum in nonmonotonic stress – strain rate relationship ($\sigma_c^*$). On the other hand, the material near the stationary wall that experiences very weak shear rate continues to age causing time dependent enhancement in $\lambda$ and $\sigma_y^*$ as a function of time. With increase in time $\sigma^*$ induced in the flow cell becomes homogeneous such that it is below $\sigma_y^*$ near the stationary wall while above $\sigma_y^*$ the moving wall. This naturally results in two prominent bands: flowing and stationary. However as shown in figure 4(a) and the inset of 4(c) the step change in $\sigma_y^*$ moves towards the moving wall causing contraction of flowing band with time. Finally in a limit of significantly high times $t^* - t_w^* > 10^4$ eventually an apparent steady state is reached, wherein flow profile may get modified extremely slowly, but the material in strationary band continues to age. In figure 4(c) we plot evolution of stress at the top plate as a function of time. It can be seen that stress decreases with time at early times $t^* - t_w^* < 1$, however slowly increases with time as material in the stationary band ages.

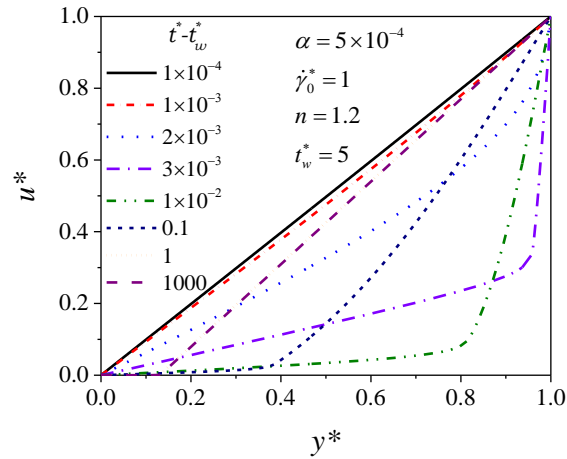

(a)

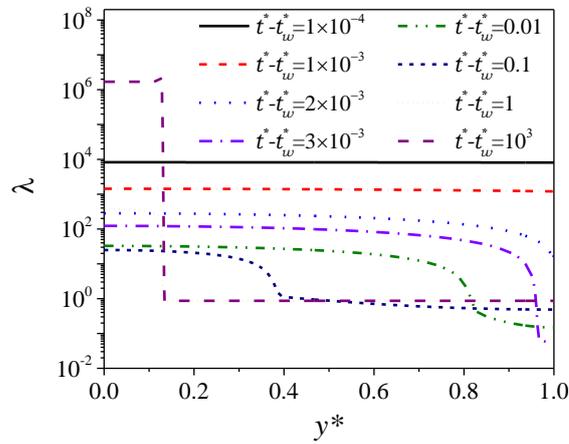

(b)

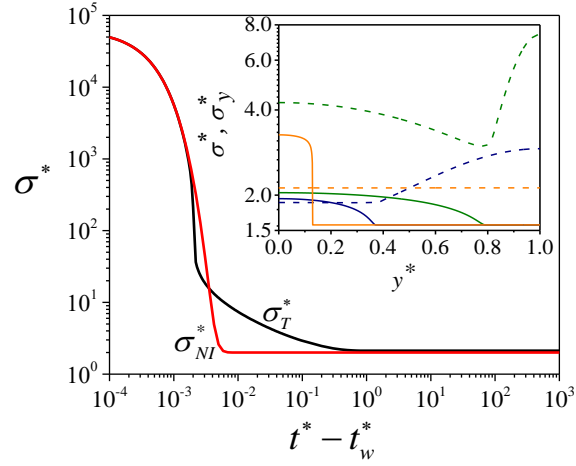

(c)

**Figure 5.** Velocity profiles (a), $\lambda$ profiles (b), and $\sigma_T^*$ and (c) at different $t^* - t_w^*$, where $t_w^*$ =5 for $n = 1.2$, $\alpha = 5 \times 10^{-4}$ under application of $\dot{\gamma}_0^* = 0.2 < \dot{\gamma}_c^*$. As expected the velocity profile in a limit of small $t^* - t_w^*$ is near homogeneous. With increase in $t^* - t_w^*$ the profile deviates from the homogeneous profile leading to transient shear banding. With further increase in $t^* - t_w^*$ the width of flowing band increases eventually leading to apparent steady state banding. In figure (c), both $\sigma_T^*$ and $\sigma_{NI}^*$ show decrease with $t^* - t_w^*$ (or equivalently the strain) eventually leading to a time invariant value. Inset shows stress (dashed) and yield stress (solid) profiles for time $t^* - t_w^* = 0.01$ (green), 0.1 (blue) and 1 (orange).

Figures 5 shows the effect of increasing $\dot{\gamma}_0^*$ to 1, with $n = 1.2$, $\alpha = 5 \times 10^{-4}$ and $t_w^* = 5$. In figures 5(a), and (b) we respectively plot velocity and $\lambda$, while in the inset of 5(c) we plot $\sigma^*$ and $\sigma_y^*$ as a function of $y^*$. Owing to the greater age of the material, both $\lambda$ (and therefore higher viscosity) and $\sigma_y^*$ are significantly higher when step shear rate is applied. However $\sigma^*$ induced in the material is even higher throughout the flow domain (not shown), which leads to the velocity profile that is near homogeneous at $t^* - t_w^* = 10^{-3}$ with a slightly higher velocity gradient closer to the moving wall. As a result, at $t^* - t_w^* = 10^{-2}$, the rejuvenation near the moving wall causes steep decrease in $\lambda$ and $\sigma_y^*$ in the region $y^* \approx 0.75 - 0.85$. At this time, as shown in the inset of figure 4(c), though $\sigma^*$ induced in the material is above $\sigma_y^*$ throughout the flow field, the difference is greater near the moving wall compared to the stationary wall. Cumulatively these effects lead to two clear shear bands as shown in figure 5(a). As apparent from $\lambda$ and $\sigma_y^*$ profile evolutions, with increase in time, the rejuvenation progresses towards the stationary wall on one hand, while the overall $\sigma^*$ induced in the material decreases on the other. Therefore, at some point, $\sigma^*$ near the stationary plate decreases below $\sigma_y^*$ as shown in figure 4(c) and eventually the point of cross over between $\sigma^*$ and $\sigma_y^*$ moves towards the stationary wall leading to expansion of the flowing band. As a result, the flowing band expands with time, but since the region closer to the stationary wall continues to age, its growth eventually gets arrested. This leads to an apparent steady state shear banding. In figure 5(c) evolution of stress at the top plate is plotted as a function of time. It can be seen that stress decreases with time at early times and eventually reaches a steady value. In principle, similar to that reported in figure 4(c) stress at the top plate should have increased with time in a limit of large $t^* - t_w^*$ due to aging of the stationary band. However, owing to small width of the stationary band for this case, the increase in stress does not seem to be noticeable.

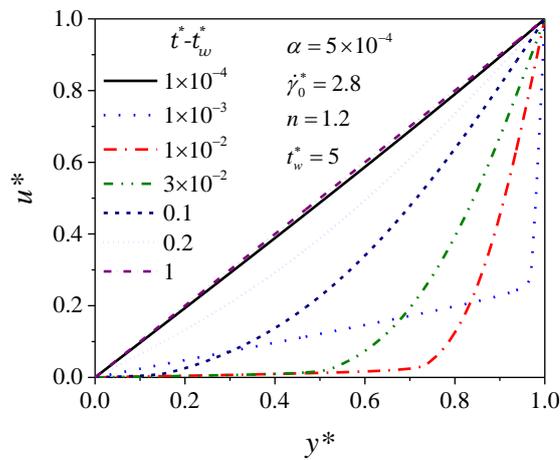

(a)

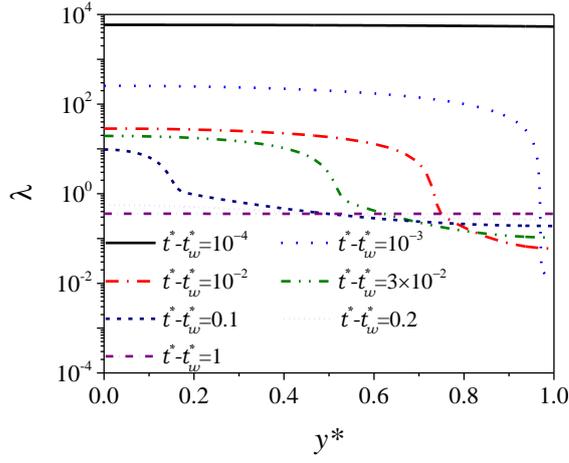

(b)

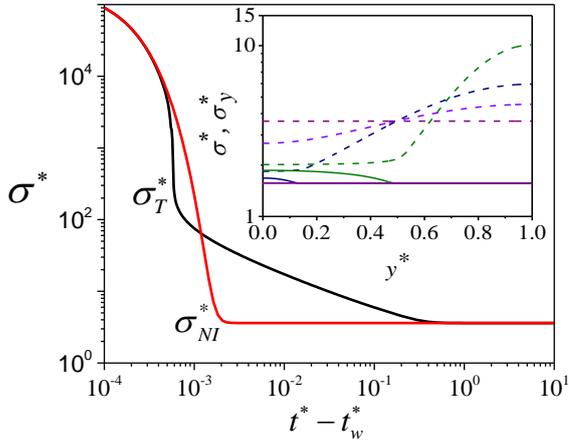

(c)

**Figure 6.** Velocity profiles (a), $\lambda$ profiles (b), and $\sigma_T^*$ and $\sigma_{NI}^*$ (c) at different $t^* - t_w^*$, where $t_w^* = 5$ for $n = 1.2$, $\alpha = 5 \times 10^{-4}$ under application of $\dot{\gamma}_0^* = 2.8 \gg \dot{\gamma}_c^*$. The velocity profile in a limit of small $t^* - t_w^*$ is near homogeneous. With increase in $t^* - t_w^*$ the profile deviates from the homogeneous profile leading to transient shear banding. With further increase in $t^* - t_w^*$, however, the flow profile eventually becomes homogeneous. In figure (c), both $\sigma_T^*$ and $\sigma_{NI}^*$ show decrease with $t^* - t_w^*$ (or equivalently the strain) eventually leading to a time invariant value. Inset shows stress (dashed) and yield stress (solid) profiles for time $t^* - t_w^* = 0.03$ (green), 0.1 (blue), 0.2 (purple) and 1 (violet).

The case of $\dot{\gamma}_0^* = 0.2$ and $\dot{\gamma}_0^* = 1$ belong to either side of the minimum associated with steady state stress-strain curve shown in figure 1. However in the both cases an 'apparent' steady state shear banding is observed. Purely based on the steady-state flow curve, one might have expected a banded state for $\dot{\gamma}_0^* = 0.2$, and a homogeneous flow state for $\dot{\gamma}_0^* = 1$. However, even for $\dot{\gamma}_0^* = 1$, due to a competetion between aging and rejuvenation during the transient evolution, we observe an apparent steady state band, and not the homogeneous state. This is solely due to the presence of inertial effects in the system. We now analyze behavior at higher shear rate of $\dot{\gamma}_0^* = 2.8$ applied at $t_w^* = 5$ for the same system ($n = 1.2$ and $\alpha = 5 \times 10^{-4}$), whose $u^*$ and $\lambda$ profiles at various times are plotted respectively in figures 6(a) and (b). In the inset of figure 6(c) we plot $\sigma^*$ and $\sigma_y^*$ as a function of $y^*$ for selective $t^* - t_w^*$. It can be seen that similar to that observed for lower shear rates, the velocity profile at $t^* - t_w^* = 10^{-4}$ is near homogeneous and slight variation in velocity gradient in this profile (greater gradient towards the moving wall) leads to greater rejuvenation in the vicinity of moving wall as suggested from the $\lambda$ and $\sigma_y^*$ profile. This leads to transient bands of slow and fast flowing material. As time progresses, the rejuvenation proceeds towards the stationary wall, and since $\sigma^*$ induced in the material always remains greater than $\sigma_y^*$ throughout the flow and time domain, the width of slow-flowing band diminishes leading to homogeneous steady state profile. As expected, in the limit of steady state flow, $\lambda$ and $\sigma_y^*$ profile also become homogeneous over the entire flow field. Evolution of $\sigma_T^*$ plotted in figure 6(c) shows a monotonic decrease that plateaus to a constant value in a limit of steady state.

The behavior reported in figures 4, 5 and 6 show striking similarity with the experimental observations of Martin and Hu [34]. For aqueous suspension of Laponite they propose the constitutive nature of the flow curve to be non-monotonic (qualitatively similar to that reported in figure 2 for $\mu > 1$) and experimentally estimate the value of critical shear rate associated with the minimun $\dot{\gamma}_c$. Their experiments with imposed shear rates $\dot{\gamma}_0$ less than $\dot{\gamma}_c$ lead to apparent steady state shear banding by means of flowing band shrinking as shown figure 4. For $\dot{\gamma}_0$ slighlty greater than $\dot{\gamma}_c$ they also report apparent steady state shear banding via flowing band expanding as predicted in figure 5. Moreoever for $\dot{\gamma}_0$ sufficiently greater than $\dot{\gamma}_c$ they observe transient shear bands, wherein the flowing band expands and fills the entire gap leading to a steady state with homogeneous flow field repored as in the figure 6. Furthermore, in all the cases, stress (this is equivalent to $\sigma_T^*$ obtained from our simulations) is observed to decrease with time as observed experimentally. Remarkably, the experiments with imposed $\dot{\gamma}_0 < \dot{\gamma}_c$ show slight increase in stress with time even after apparent steady state has set in as also predicted by the simulations as shown in figure 4(c).

In order to obtain further insight into how the shear rate in the flowing band ($\dot{\gamma}_{FB}$) changes with the imposed shear rate ($\dot{\gamma}_0$) in a limit of steady state (apparent or real), we plot experimental data of $\dot{\gamma}_{FB}/\dot{\gamma}_c$ as a function of $\dot{\gamma}_0/\dot{\gamma}_c$ extracted from Martin and Hu [34] in the top inset of fgure 7. It can be seen that for $\dot{\gamma}_0$ sufficiently greater than $\dot{\gamma}_c$ flow field becomes homogeneous ($\dot{\gamma}_{FB} = \dot{\gamma}_0$) at sufficiently long times. However at the lower shear rates the system shows apparent steady state shear banding ($\dot{\gamma}_{FB} > \dot{\gamma}_0$) even though $\dot{\gamma}_0 > \dot{\gamma}_c$. Furthermore, for the shear rates $\dot{\gamma}_0 < \dot{\gamma}_c$ as well apparent steady state shear banding is observed with $\dot{\gamma}_{FB} \approx \dot{\gamma}_c$. In order to assess how the simulations compare with the experimental observations we plot shear rate in the flowing band ($\dot{\gamma}_{FB}^*$) with the imposed shear rate ($\dot{\gamma}_0^*$) in a limit of apparent or real steady state for $t_w^* = 0$ and 5 and for five values of $\alpha$ ranging from $5\times10^{-5}$ and 5. The present work clearly shows that shear banding in physically aging systems originates primarily due to interplay between dominance of rejuvenation near stationary wall and that of aging near the moving wall. The location of interface between continuously rejuvenating region and continuously aging region, which describes apparent steady state shear banding, depends on parameter $\alpha$ and the imposed strain rate. The greater the value of $\alpha$ is, the faster is the momentm diffusivity compared to the timescale asscoaited with aging (it should be noted that $\alpha$ is defined as: $\alpha = T_0\eta_0/\rho H^2$ suggesting ratio of time scale associated with aging ($T_0$) and that associated with momentum diffusivity only when $\lambda = 0$ that is immediately after rejuvenation. However for times $t^* > 0$, the real ratio of timescale of aging to that of momentum diffusivity is expected to be much larger than the intial value of the same given by $\alpha$). It can be seen that irrespective of value of $\alpha$, for large $\dot{\gamma}_0^*$, the homogeneous steady state is always achieved leading to $\dot{\gamma}_{FB}^* = \dot{\gamma}_0^*$. However as $\dot{\gamma}_0^*$ decreases, $\dot{\gamma}_{FB}^*$ bifurcates from the line representing $\dot{\gamma}_{FB}^* = \dot{\gamma}_0^*$ suggesting apparent steady state shear banding. With increase in $\alpha$, the bifurcation from $\dot{\gamma}_{FB}^* = \dot{\gamma}_0^*$ occurs at a smaller value of $\dot{\gamma}_0^*$. For $\alpha = 5\times10^{-5}$, after the bifurcation from $\dot{\gamma}_{FB}^* = \dot{\gamma}_0^*$, $\dot{\gamma}_{FB}^*$ decreases weakly with decrease in $\dot{\gamma}_0^*$ while for $\alpha = 5 \times 10^{-4}$, $\dot{\gamma}_{FB}^*$ can be seen to be reaching a plateau soon after the bifurcation. In both the cases the change in $\dot{\gamma}_{FB}^*$ after the bifurcation is gradual suggesting that the width of the flowing band (The lever rule suggests that the fractional thickness of flowing band: $h_{FB} = H_{FB}/H$ is related to $\dot{\gamma}_{FB}^*$, as $h_{FB} = \dot{\gamma}_0^*/\dot{\gamma}_{FB}^*$. As a result, higher a value of $\dot{\gamma}_{FB}^*$ is compared to $\dot{\gamma}_0^*$, smaller is $h_{FB}$) decreases steadily with decrease in $\dot{\gamma}_0^*$. However, for the higher values of $\alpha$, interestingly $\dot{\gamma}_{FB}^*$ is observed to jump to a high value after the bifurcation indicating sudden change in the width of the flowing band upon small decrease in $\dot{\gamma}_0^*$. Nonetheless, wherever the apparent steady state shear banding is observed, $\dot{\gamma}_{FB}^*$ is observed to decrease with increase in $\alpha$, irrespective of value of $\dot{\gamma}_0^*$.

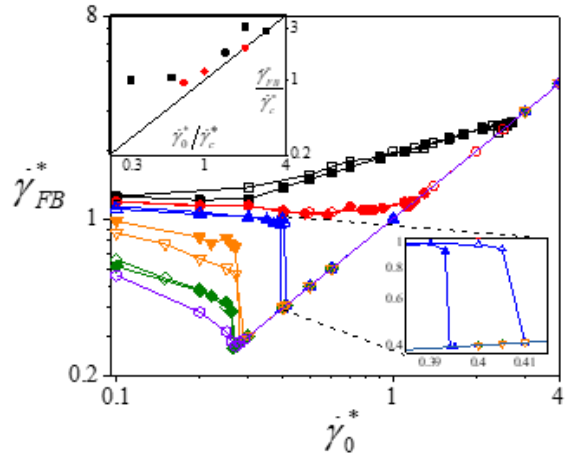

**Figure 7.** The shear rate in the flowing band ($\dot{\gamma}_{FB}^*$) is plotted against the imposed shear rate ($\dot{\gamma}_0^*$) for $n$ = 1.2 and, and when step shear rate is applied at $t_w^*$ = 0 (closed symbol) and $t_w^*$ = 5 (open symbol) for $\alpha$ = $5 \times 10^{-5}$ (squares), $5 \times 10^{-4}$ (circles), $5 \times 10^{-3}$ (up triangles), 0.05 (down triangles), 0.5 (diamonds), 5 (hexagon). The top inset shows experimental values of Martin and Hu [34], for couette (square) and cone and plate (circle) geometry.

The simulation results plotted in figure 7 appear to show that waiting time has little effect on the apparent steady state behavior. However a closer look indeed suggests that the data associated with $t_w^*$ = 0 and 5 are separate, at least over a certain domains of $\dot{\gamma}_0^*$ and $\alpha$. In the lower inset of figure 7 we show enlarged region around $\dot{\gamma}_0^*$ =0.4, which shows that for $\alpha$ =$5 \times 10^{-3}$, the data associated with $t_w^*$ = 0 bifucates from $\dot{\gamma}_{FB}^* = \dot{\gamma}_0^*$ line at $\dot{\gamma}_0^* \approx 0.394$ while that of for $t_w^* = 5$ at $\dot{\gamma}_0^* \approx 0.41$. In figure 8(a) we plot an evolution of the velocity profile for $\alpha = 5 \times 10^{-3}$ and $t_w^* = 0.1$ for $\dot{\gamma}_0^* = 0.4$ while in figure 8(b) we plot the velocity profiles for the same values of $\alpha$ and $\dot{\gamma}_0^*$ but for $t_w^* = 5$. It can be seen that, as expected, for $t_w^* = 0.1$, the velocity profile becomes homogeneous in the limit of steady state without showing any transient banding. On the other hand, for $t_w^* = 5$ the initially near-homogeneous profile develops to show transient as well as apparent steady state shear banding. Interestingly Kurokawa and coworkers [32] also report a qualitaively similar situation wherein velocity profiles for step shear rate applied at lower waiting time evole to a homogeneous steady state (although after showing transient shear banding) while for the step shear rate applied at large waiting times apparent steady state shear banding is observed.

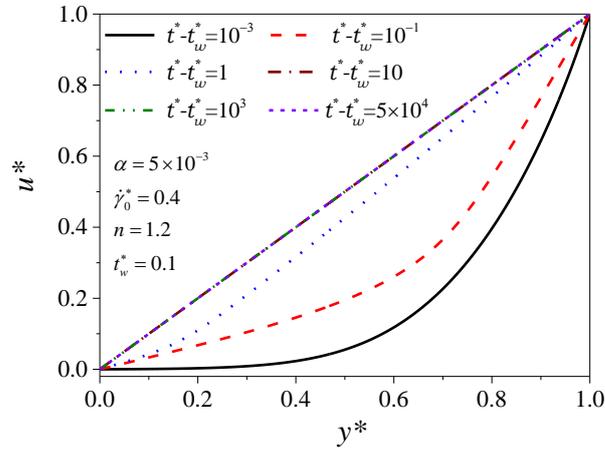

(a)

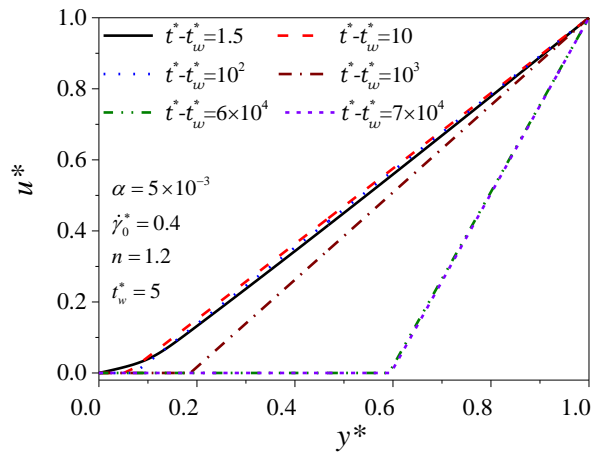

(b)

**Figure 8.** Velocity profiles for $n = 1.2$, $\alpha = 5 \times 10^{-3}$ when subjected to step shear rate $\dot{\gamma}_0^* = 0.4$ at $t_w^* = 0.1$ (a) and 5 (b). It can be seen that application of step shear rate immediately or very short time after rejuvenation does not induce shear banding while application of the same at large times does leads to apparent steady state shear banding.

The case of $n=1$ in the present formulation leads to stress plateau in a limit of low shear rates. According to Joshi [33] this scenario occurs where system shows linear dependence of relxation time on waiting time, which is also implied by equation (4) wherein viscosity shows linear increase on time under quiscent conditions. Similarly for $n>1$ dependence of relaxation time (or vicosity) on waiitng time is faster than linear in absence of flow that leads to non-monotonic steady state stress – shear rate relation. The present work suggests transient shear banding to be originating from ability of material near the stationary wall to age rapidly before rejuvenation sets in. We therefore feel that this dynamics of material wherein relaxation time grows stronger than linear is responsible for it showing transient shear banding.

<u>Importance of inertia in transient banding</u>

Most of the work [35,44-46,53] carried out on simulating shear banding in soft glassy materials ignores inertia, which implies that $\underset{\sim}{\nabla}\cdot\underset{\approx}{\sigma}=0$. As a result, in order to have any inhomogeneity in the flow field there must either be inhomogeneity in shear stress and/or the parameters of the constitutive relation. Regarding the former, shear stress is constant across the flow field for planer Couette geometry (shown in figure 1). In practice the experimental investigations on the shear banding have been carried out either in concentric cylinder or cone and plate geometry. For concentric cylinder system there is a weak variation of stress in the gradient direction for the dimensions usually used for the experiments. However, in the cone and plate geometry the variation in the gradient direction is negligible for the typical dimensions employed in the experiments. Moreover, the parameters such as viscosity, modulus, etc., of a constitutive relation that depend on fluidity can become inhomogeneous only if the fluidity evolution equation has a term that shows explicit dependnce on the gradient direction (such as diffusive term used in equation (1)). When stress is homogeneous across the flow field, inhomogeniety in fluidity across the flow field is needed at the point of application of step shear rate, even in the presence of a diffusive term, in order to initiate the banding process in the simulations.

A careful examination of the work of Moorcroft *et al.* [44] indicates that transient banding in their formulation requires the following key ingredients: (1) The presence of a diffusive term in the equation governing the dynamics of the relaxation time, (2) the presence of a viscoelastic stress for the glassy degrees of freedom, and (3) an initial inhomogeneity in the stress initial condition. In contrast, in the present study, we show that transient banding, especially in constitutive relations with non-monotonic flow curves (in terms of stress vs strain rate), can happen even in the absence of the above ingredients required in Moorcroft *et al.* [44]. The only ingredients required for transient banding in the present model are (1) inclusion of inertial effects that are inevitably present in any start-up flow experiment, and (2) non-monotonicity in the stress-strain rate flow curve. In the earlier studies [35,44], the diffusive term in the structure evolution plays an important role in the formation of the transient bands. However, in this study, we neglected this diffusive term, but instead included fluid inertia, which naturally leads to the diffusive transport of momentum. Thus, the transient and apparent steady-state bands in the present work are formed due to momentum diffusion instead of the diffusive term in the structure evolution equation. Further, Moorcroft *et al.* [44] also had to impose an initial inhomogeneous stress distribution in their simulations in order to achieve any banding. Rogers *et al.* [35], on the other hand, incorporate a noise term in the evolution equation that dpends on the gradient direction. In the present work, the momentum diffusion process (at nonzero inertia) naturally leads to a stress inhomogeneity and hence precludes the need to impose it explicitly as an initial condition. We therefore feel that since momentum diffusion is always important in any transient start-up dynamics, it may not be ignored in any description of transient banding.

In the work of Alava and co-workers [50,51], the steady-state flow curve is monotonic, with zero yield stress. In marked contrast, the present work considers a more general constitutive model that is thixotropic in nature, which includes as its special cases (i) monotonic stress-strain rate curve, (ii) an yield stress fluid, and (iii) a nonmonotonic stress-strain rate curve. Thus, when $n<1$ the steady state flow curve of our model shares some qualitative similarities to the model used by Alava and co-workers [50,51]. However, we do not observe distinct transient shear banding for $n<1$ in our model, in contrast to the results of Alava and co-workers [50,51], although there is some sharp velocity variation in our study as well at early times for $n=1$. This variation however, is not as sharp and well-defined as the banding that takes place in our model for $n=1.2$.

Our results thus show that when the momentum diffusion time scale is finite, transient and apparent steady state banding happen rather naturally arising from a competition between rejuvenation due to momentum diffusion and aging. This happens even in the absence of any viscoelastic effects in the constitutive model and in the absence of diffusive dynamics in the fluidity model.

Assessing the criteria for onset of transient banding

Our results also point to some important counter examples for the criteria outlined in the earlier studies of Fielding and co-workers [44,45,53] for the initiation of transient shear banding. We confine our discussion only to the protocol analyzed in the present study, viz., a step increase in shear rate in the flow. The other important case of step increase in stress discussed by Moorcroft and Fielding [45] will be addressed in a subsequent work.

Using a linear stability analysis on the instantaneous velocity profile that arises due to the step change in shear rate, Moorcroft *et al.* [44,45] concluded that the necessary condition for transient banding is the non-monotonicity in the stress-time (or strain) curve. Indeed, their analysis shows that the homogeneous flow is unstable in the decreasing part of the stress-strain curve. However, their analysis pertains to the situation wherein effect of inertia is completely ignored. Therefore we also estimate the stress induced in material assuming no inertia ($\sigma_{NI}^*$, which is obtained by solving only the constitutive equation by assuming $\nabla \cdot \underset{\sim}{\sigma} = 0$ or $\alpha = \infty$, the subscript NI represents 'no inertia') for all the cases studied above and plot the same along with $\sigma_T^*$ that has been obtained by considering inertia. The time evolution of $\sigma_{NI}^*$ for the present model is given by:

$$\sigma_{NI}^* = \dot{\gamma}_0^* \left[ 1 + \left\{ \frac{1}{\dot{\gamma}_0^*} \left( 1 - \left( 1 - \frac{\dot{\gamma}_0^*}{\alpha} t_w^* \right) \exp\left( -\frac{\dot{\gamma}_0^*}{\alpha} t^* \right) \right) \right\}^n \right] \tag{16}$$

In our study, we have clearly found instances (Figure 3, $n=1$, $t_w^* = 0.5$), where $\sigma_T^*$ as well as $\sigma_{NI}^*$ indeed decrease with time (after the step change in shear rate), but yet the system does not show any transient banding while approaching the homogeneous steady state. In figure 9(a), we plot evolution of velocity profiles for $n=1$, $t_w^* = 0$, for which $\sigma_T^*$ decreases with time while $\sigma_{NI}^*$ increases with time. In this case as well no transient as well as steady state shear banding has been observed. Furthermore, as reported in figures 4(c), 5(c) and 6(c) wherein $n=1.2$ and $t_w^* = 5$ for different $\dot{\gamma}_0^*$ on the either side of $\dot{\gamma}_c^*$, $\sigma_T^*$ as well as $\sigma_{NI}^*$ both decrease with time and transient shear banding is observed in all the cases. In figure 9(b), 9(c) and 9(d), we plot evolution of velocity profiles for $n=1.2$ and $t_w^* = 0$ while in the insets of the respective figures we plot corresponding time evolutions of $\sigma_T^*$ and $\sigma_{NI}^*$ for different imposed strain rates. The behaviour of $\sigma_T^*$ as well as $\sigma_{NI}^*$ qualitatively similar for all the cases, i.e. $\sigma_T^*$ decreases with time while $\sigma_{NI}^*$ monotonically increases with time. We observe that for $n = 1.2$ and for $\dot{\gamma}_0^* = 0.2$ and 1.2, there is both transient and apparent steady-state banding. When $\dot{\gamma}_0^*$ is increased to 2.8, however there is very little transient banding, and no steady-state banding. The combined analysis of all the figures plotted in figure 9 show that the four cases exhibit very different banding behaviour despite the stress profiles being qualitatively identical.

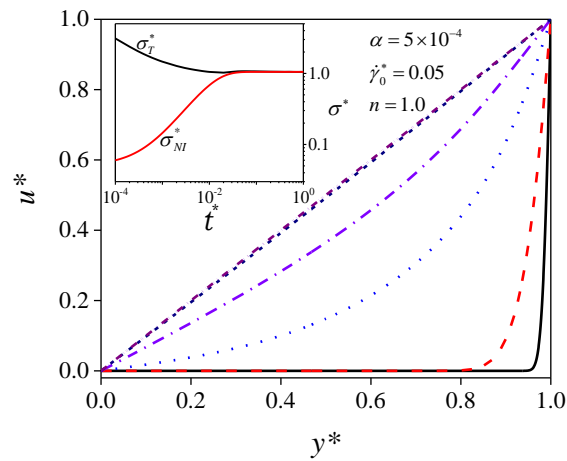

(a)

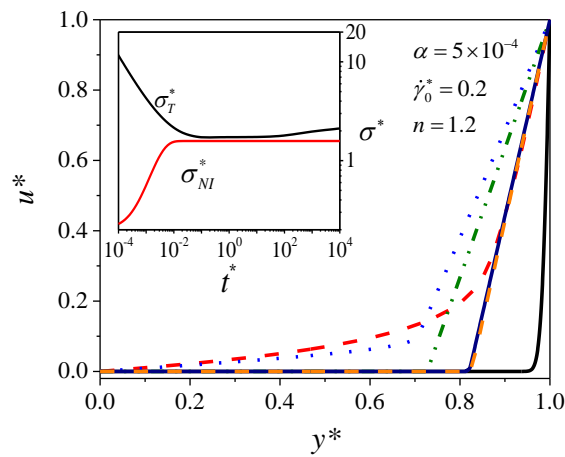

(b)

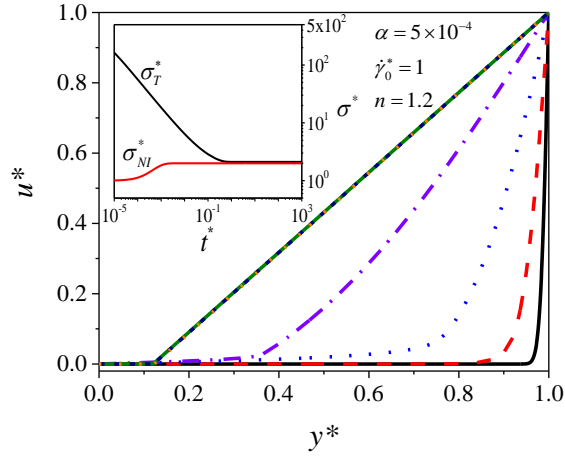

(c)

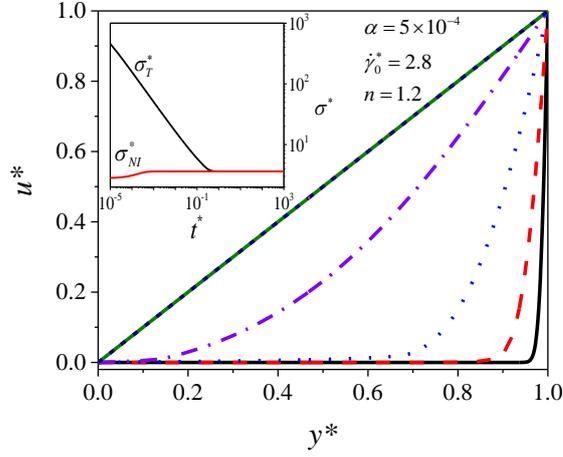

(d)

**Figure 9.** The main figure shows the time evolution of velocity profiles while the inset shows $\sigma_T^*$ and $\sigma_{NI}^*$ at different $t^* - t_w^*$, when step shear rate is applied at $t_w^* = 0$ for (a) $n = 1$, $\alpha = 5 \times 10^{-4}$ and $\dot{\gamma}_0^* = 0.05$ (from right to left, $t^* = 10^{-4}, 10^{-3}, 10^{-2}, 10^{-1}, 1, 10, 10^2$), (b) $n = 1.2$, $\alpha = 5 \times 10^{-4}$ and $\dot{\gamma}_0^* = 0.2$ ($t^* = 10^{-4}$ (black), $10^{-2}$ (red), $10^{-1}$ (blue), $10^2$ (green), $7 \times 10^3$ (navy), $10^4$ (orange)). (c) $n = 1.2$, $\alpha = 5 \times 10^{-4}$ and $\dot{\gamma}_0^* = 1$ (from right to left, $t^* = 10^{-4}, 10^{-3}, 10^{-2}, 10^{-1}, 1, 10, 10^3$). (d) $n = 1.2$, $\alpha = 5 \times 10^{-4}$ and $\dot{\gamma}_0^* = 2.8$ (from right to left, $t^* = 10^{-4}, 10^{-3}, 10^{-2}, 10^{-1}, 1, 10, 10^3$). Interestingly all the cases $\sigma_T^*$ decreases with time while $\sigma_{NI}^*$ increases with time. On the other hand, the case of $n = 1$ does not show any kind of transient shear banding, the case of $n = 1.2$ and $\dot{\gamma}_0^* = 0.2$ and 1 show transient as well as apparent steady state shear banding, while $n = 1.2$ and $\dot{\gamma}_0^* = 2.8$ shows only transient shear banding while no steady state shear banding.

The stress evolutions described for all the cases studied in the present work clearly suggest that irrespective of any combination of parameters $\sigma_T^*$ always decreases with time as system approaches the steady state and when system shows or does not show the transient and/or homogeneous steady state or apparent steady state shear banding. This result is not surprising as even for the simplest case of transient start up of planar Couette flow of a Newtonian fluid, the stress-strain (or time) relation has always a negative slope wherein neither transient nor steady state shear banding is observed [54]. Evolution of $\sigma_{NI}^*$ on the other hand shows monotonic increase for $t_w^* = 0$ while monotonic decrease for $t_w^* > 0$. However, as shown in this study neither of the cases gurantees transient as well as apparent steady state shear banding. Furthermore, as shown in figures 3(a) initial state of homogeneous flow (deformation) also does not ensure transient and apparent steady state shear banding.

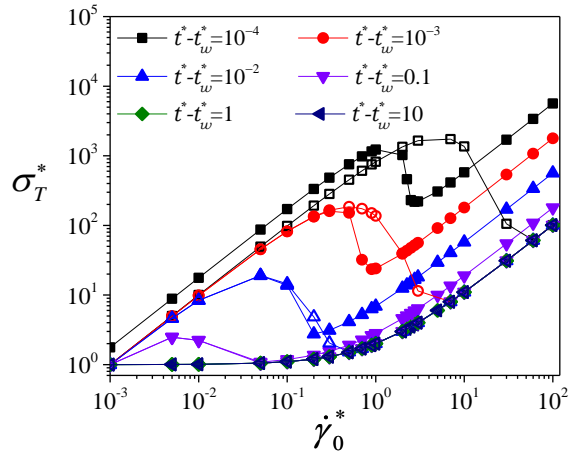

**Figure 10.** Stress at the top plate ($\sigma^*$ at $y^* = 1$) is plotted as a function of step shear rate $\dot{\gamma}_0^*$ applied at $t_w^* = 0.5$ for $n = 1$ and $\alpha = 5 \times 10^{-4}$. (Open symbol: Non inertial system, Closed symbol: Inertial system). It is important to note that while the above relationship does show non-monotonicity, the corresponding velocity profile does not show any kind of shear banding.

The special case of $n = 1$ in our work corresponds to a constant yield stress behaviour. As shown in figure 10, for $n = 1$ even our model shows a non-monotonic behaviour in terms of stress vs imposed strain rate obtained at different constant values of $t^* - t_w^*$, after an application of a step change in shear rate. This figure is identical in construction (for the present model) to figure 2(a) of Moorcroft *et al.* [44] for their model. In their work, by carrying out instantaneous stability analysis Moorcroft *et al.* [44], propose that such non-monotonicity results in transient banding. However, our numerical simulations for this case do not show any transient banding for $n=1$. This suggests that the presence of non-monotonic behaviour in the stress-strain diagram or in the stress-strain rate (at constant $t^* - t_w^*$) does not have any corelation to the transient banding.

This discussion demonstrates that the formation of transient bands in soft-glassy systems is a more complex phenomenon compared to the prevalent theoretical understanding of the same. Inclusion of inertial effects can be very crucial in the modelling of transient bands, as in any realistic system involving transient dynamics, inertia is expected to play a subtle but important role. This is unlike the prediction of steady-state banding in time-independent fluids, where the path to the eventual steady state is perhaps irrelevant to the eventual steady state reached.

Therefore, the central inference from our work is that when transient dynamics is considered, inertial effects naturally become important, and more so in the case of time-dependent soft-glassy materials. Here, the stress inhomogeneities created at early times by the sudden imposition of shear rate are inextricably linked to the structure evolution in the system, and these two effects are tightly coupled. It thus seems reasonable to speculate that it is the presence of inertia in the present model that results in disagreement with the criteria proposed by Fielding [53], which were based upon an instantaneous stability analysis in the limit of no inertia. Therefore, unlike apparent steady-state banding, where a non-monotonicity in the stress-shear rate curve is necessary, the mechanisms for the development of transient bands, especially in soft glassy materials, could be more complicated as to be subsumed into a single criterion for transient banding.

It should be noted that our model is inelastic in nature, and for non-monotonic flow curves is able to qualitatively capture the experimental observations for the soft glassy materials. However for $n=1$ our results do not exhibit any kind of banding (transient or otherwise). There could still be a posisbility that for $n=1$ system may show transient banding if elasticity is considered in a constitutive model along with inertial effects in momentum balance equation. However, the simultaneous consideration of elasticity and inertia is known to cause formation of elasto-inertial waves [48,49] leading to transient banding even in time – independent models [47]. It is therefore conceivable that even in aging fluids with monotonic flow curves the inclusion of elasticity and inertia may lead to transient banding.

## IV. Conslusions:

In this work we studied transient as well as apparent steady state shear banding behavior of a fluidity model subjected to step shear rate in rectilinear Couette flow by accounting for fliud inertia during the transient evolution. The fluidity model used in the present work leads to monotonic as well as non-monotonic steady state stress – shear rate flow curves depending upon the choice of a parameter, and captures the rheological features of time-dependent aging soft glassy materials well. We observe that when flow curve is monotonic the velocity profile evolves to final homogeneous steady state without showing transient shear banding. On the other hand when flow curve is non-monotonic in nature, the system shows transient and/or apparent sheady state shear banding depending upon magnitude of employed shear rate and timescale for momentum diffusivity. Our work suggests that spacially varying rate of aging altered by that of rejuvenation through diffusion of momentum is a vital aspect of shear banding. Very importantly, the results from our simulations show very close qualitative agreement with some of the experimental observations reported for the aging soft glassy materials. Considering that the present model is a minimal model to capture only the essential aspects of aging dynamics, this close agreement is indeed encouraging. We also observe that negative slope of the stress – strain (or time) relation during the transient and/or presence of near homogeneous velocity profile immediately after application of step shear rate have no direct correlation to the transient as well as steady state shear banding. Earlier studies have neglected inertia, but included elasticity, diffusion in the structrure evolution as essential ingredients to achieve transient banding. In marked contrast, our work suggests that inclusion of inertia in the transient dynamics can also be very important in the relastic description of shear banding behavior (transient or otherwise) of time-dependent fluids.

**Acknowledgment:** The financial support from the department of Atomic Energy- Science Research Council, Government of India is greatly acknowledged.

## References:

[1]	P. Coussot, *Rheometry of Pastes, Suspensions, and Granular Materials* (John Wiley & Sons, Inc., Hoboken, 2005).

[2]     J. Mewis and N. J. Wagner, *Colloidal Suspension Rheology* (Cambridge University Press, Cambridge, 2012).
[3]     S. M. Fielding, P. Sollich, and M. E. Cates, "Aging and rheology in soft materials" Journal of Rheology **44**, 323 (2000).
[4]     P. Coussot, "Rheophysics of pastes: A review of microscopic modelling approaches" Soft Matter **3**, 528 (2007).
[5]     I. M. Hodge, "Physical aging in polymer glasses" Science **267**, 1945 (1995).
[6]     R. Bandyopadhyay, D. Liang, J. L. Harden, and R. L. Leheny, "Slow dynamics, aging, and glassy rheology in soft and living matter" Solid State Communications **139**, 589 (2006).
[7]     G. B. McKenna, T. Narita, and F. Lequeux, "Soft colloidal matter: A phenomenological comparison of the aging and mechanical responses with those of molecular glasses" Journal of Rheology **53**, 489 (2009).
[8]     Y. M. Joshi, "Dynamics of Colloidal Glasses and Gels" Annual Review of Chemical and Biomolecular Engineering **5**, 181 (2014).
[9]     M. Cloitre, R. Borrega, and L. Leibler, "Rheological aging and rejuvenation in microgel pastes" Physical Review Letters **85**, 4819 (2000).
[10]    D. Bonn, S. Tanase, B. Abou, H. Tanaka, and J. Meunier, "Laponite: Aging and shear rejuvenation of a colloidal glass" Physical Review Letters **89**, 015701 (2002).
[11]    V. Viasnoff and F. Lequeux, "Rejuvenation and Overaging in a Colloidal Glass under Shear" Physical Review Letters **89**, 065701 (2002).
[12]    C. Christopoulou, G. Petekidis, B. Erwin, M. Cloitre, and D. Vlassopoulos, "Ageing and yield behaviour in model soft colloidal glasses" Philosophical Transactions of the Royal Society A: Mathematical, Physical and Engineering Sciences **367**, 5051 (2009).
[13]    P. Coussot, Q. D. Nguyen, H. T. Huynh, and D. Bonn, "Viscosity bifurcation in thixotropic, yielding fluids" Journal of Rheology **46**, 573 (2002).
[14]    D. Bonn, M. M. Denn, L. Berthier, T. Divoux, and S. Manneville, "Yield stress materials in soft condensed matter" Reviews of Modern Physics **89**, 035005 (2017).
[15]    T. Divoux, M. A. Fardin, S. Manneville, and S. Lerouge, "Shear Banding of Complex Fluids" Annual Review of Fluid Mechanics **48**, 81 (2016).
[16]    M. Kaushal and Y. M. Joshi, "Tailoring relaxation time spectrum in soft glassy materials" Journal of Chemical Physics **139** (2013).
[17]    M. Kaushal and Y. M. Joshi, "Analyzing aging under oscillatory strain field through the soft glassy rheology model" The Journal of Chemical Physics **144**, 244504 (2016).
[18]    D. Vlassopoulos and M. Cloitre, "Tunable rheology of dense soft deformable colloids" Current Opinion in Colloid & Interface Science **19**, 561 (2014).
[19]    D. Bonn and M. M. Denn, "Yield Stress Fluids Slowly Yield to Analysis" Science **324**, 1401 (2009).
[20]    A. S. Negi and C. O. Osuji, "Time-resolved viscoelastic properties during structural arrest and aging of a colloidal glass" Physical Review E **82**, 031404 (2010).
[21]    P. Coussot, Q. D. Nguyen, H. T. Huynh, and D. Bonn, "Avalanche behavior in yield stress fluids" Phys. Rev. Lett. **88**, 1755011 (2002).
[22]    J. Sprakel, S. B. Lindström, T. E. Kodger, and D. A. Weitz, "Stress Enhancement in the Delayed Yielding of Colloidal Gels" Physical Review Letters **106**, 248303 (2011).
[23]    B. Baldewa and Y. M. Joshi, "Delayed yielding in creep, time-stress superposition and effective time theory for a soft glass" Soft Matter **8**, 789 (2012).
[24]    A. Shukla and Y. M. Joshi, "Ageing under oscillatory stress: Role of energy barrier distribution in thixotropic materials" Chemical Engineering Science **64**, 4668 (2009).
[25]    Y. M. Joshi, A. Shahin, and M. E. Cates, "Delayed solidification of soft glasses: New experiments, and a theoretical challenge" Faraday Discussions **158**, 313 (2012).
[26]    A. Negi and C. Osuji, "Dynamics of a colloidal glass during stress-mediated structural arrest" EPL (Europhysics Letters) **90**, 28003 (2010).

[27]     V. Viasnoff, S. Jurine, and F. Lequeux, "How are colloidal suspensions that age rejuvenated by strain application?" Faraday Discussions **123**, 253 (2003).
[28]     R. B. Bird, R. C. Armstrong, and O. Hassager, *Dynamics of Polymeric Liquids, Fluid Mechanics* (Wiley-Interscience, New York, 1987).
[29]     N. J. Balmforth, I. A. Frigaard, and G. Ovarlez, "Yielding to stress: Recent developments in viscoplastic fluid mechanics" Annual Review of Fluid Mechanics **46**, 121 (2014).
[30]     G. Ovarlez, S. Rodts, X. Chateau, and P. Coussot, "Phenomenology and physical origin of shear localization and shear banding in complex fluids" Rheologica Acta **48**, 831 (2009).
[31]     J. Paredes, N. Shahidzadeh-Bonn, and D. Bonn, "Shear banding in thixotropic and normal emulsions" Journal of Physics: Condensed Matter **23**, 284116 (2011).
[32]     A. Kurokawa, V. Vidal, K. Kurita, T. Divoux, and S. Manneville, "Avalanche-like fluidization of a non-Brownian particle gel" Soft Matter **11**, 9026 (2015).
[33]     Y. M. Joshi, "A model for aging under deformation field, residual stresses and strains in soft glassy materials" Soft Matter **11**, 3198 (2015).
[34]     J. D. Martin and Y. Thomas Hu, "Transient and steady-state shear banding in aging soft glassy materials" Soft Matter **8**, 6940 (2012).
[35]     S. A. Rogers, D. Vlassopoulos, and P. T. Callaghan, "Aging, Yielding, and Shear Banding in Soft Colloidal Glasses" Physical Review Letters **100**, 128304 (2008).
[36]     T. Divoux, D. Tamarii, C. Barentin, and S. Manneville, "Transient Shear Banding in a Simple Yield Stress Fluid" Physical Review Letters **104**, 208301 (2010).
[37]     T. Divoux, D. Tamarii, C. Barentin, S. Teitel, and S. Manneville, "Yielding dynamics of a Herschel-Bulkley fluid: a critical-like fluidization behaviour" Soft Matter **8**, 4151 (2012).
[38]     T. Divoux, C. Barentin, and S. Manneville, "From stress-induced fluidization processes to Herschel-Bulkley behaviour in simple yield stress fluids" Soft Matter **7**, 8409 (2011).
[39]     V. Grenard, T. Divoux, N. Taberlet, and S. Manneville, "Timescales in creep and yielding of attractive gels" Soft Matter **10**, 1555 (2014).
[40]     G. Ovarlez, S. Cohen-Addad, K. Krishan, J. Goyon, and P. Coussot, "On the existence of a simple yield stress fluid behavior" Journal of Non-Newtonian Fluid Mechanics **193**, 68 (2013).
[41]     P. Sollich, F. Lequeux, P. Hebraud, and M. E. Cates, "Rheology of soft glassy materials" Phys. Rev. Lett. **78**, 2020 (1997).
[42]     S. M. Fielding, M. E. Cates, and P. Sollich, "Shear banding, aging and noise dynamics in soft glassy materials" Soft Matter **5**, 2378 (2009).
[43]     J. Mewis and N. J. Wagner, "Thixotropy" Advances in Colloid and Interface Science **147–148**, 214 (2009).
[44]     R. L. Moorcroft, M. E. Cates, and S. M. Fielding, "Age-Dependent Transient Shear Banding in Soft Glasses" Physical Review Letters **106**, 055502 (2011).
[45]     R. L. Moorcroft and S. M. Fielding, "Criteria for Shear Banding in Time-Dependent Flows of Complex Fluids" Physical Review Letters **110**, 086001 (2013).
[46]     R. Radhakrishnan, T. Divoux, S. Manneville, and S. M. Fielding, "Understanding rheological hysteresis in soft glassy materials" Soft Matter **13**, 1834 (2017).
[47]     L. Zhou, L. P. Cook, and G. H. McKinley, "Multiple shear-banding transitions for a model of wormlike micellar solutions" SIAM Journal on Applied Mathematics **72**, 1192 (2012).
[48]     R. I. Tanner, "Note on the Rayleigh problem for a visco-elastic fluid" Zeitschrift für angewandte Mathematik und Physik ZAMP **13**, 573 (1962).
[49]     M. M. Denn and K. C. Porteous, "Elastic effects in flow of viscoelastic liquids" The Chemical Engineering Journal **2**, 280 (1971).
[50]     X. Illa, A. Puisto, A. Lehtinen, M. Mohtaschemi, and M. J. Alava, "Transient shear banding in time-dependent fluids" Physical Review E **87**, 022307 (2013).
[51]     M. Korhonen, M. Mohtaschemi, A. Puisto, X. Illa, and M. J. Alava, "Start-up inertia as an origin for heterogeneous flow" Physical Review E **95**, 022608 (2017).

[52]   R. Radhakrishnan and S. M. Fielding, "Shear Banding of Soft Glassy Materials in Large Amplitude Oscillatory Shear" Physical Review Letters **117**, 188001 (2016).
[53]   S. M. Fielding, "Triggers and signatures of shear banding in steady and time-dependent flows" Journal of Rheology **60**, 821 (2016).
[54]   L. G. Leal, *Advanced Transport Phenomena: Fluid Mechanics and Convective Transport Processes* (Cambridge University Press, Cambridge, 2007).